\documentclass[aps,prl,twocolumn,superscriptaddress,showpacs]{revtex4}
\usepackage{graphicx,epsfig}
\begin{document}
\title{Instabilities in Binary Mixtures of One-Dimensional
Quantum Degenerate Gases}
\author{M.~A. Cazalilla}
\affiliation{Donostia International Physics Center (DIPC), 
Manuel de Lardizabal 4, 20018-Donostia, Spain.}
\author{A.~F. Ho}
\affiliation{School of Physics and Astronomy,  The University of Birmingham,
Edgbaston, Birmingham B15 2TT, UK.}
\affiliation{Donostia International Physics Center (DIPC), 
Manuel de Lardizabal 4, 20018-Donostia, Spain.}
\begin{abstract}
We show that  one-dimensional binary mixtures of bosons or of a boson 
and a spin-polarized fermion  are Luttinger liquids with the following  instabilities: 
i) For different
particle densities, strong attraction between the mixture components leads to collapse, 
while strong repulsion
leads to demixing, ii) For a low-density mixture of two gases 
of impenetrable bosons (or a spin-polarized fermion
and an impenetrable boson) of  equal densities,  the system develops a gap and exhibits 
enhanced pairing fluctuations when there is attraction between the 
components. In the boson-fermion mixture, the pairing
fluctuations occur at finite momentum. Our conclusions apply to mixtures
both on the continuum and on optical lattices away from integer or fractional commensurability.
\end{abstract}
\maketitle

The exploration of  properties of matter at increasingly lower temperatures
and densities   has yielded many surprises. The achievement of Bose-Einstein
condensation (BEC) in dilute gases of ultracold atoms~\cite{Anglin02} is 
one beautiful  example.  Ever since the observation of BEC, one line of current 
experimental (and theoretical) studies is on
cold gases in very elongated atomic traps. In effectively  one-dimensional (1D)
systems  strong phase fluctuations can occur. These systems are interesting  
for both large and small $\gamma$ values, where $\gamma = M g/\hbar^2 
\rho_0$ is the dimensionless
coupling ratio,  $g$ is the coupling characterizing  the interaction between
the atoms in 1D, $M$  the atom mass, and $\rho_0$
the particle density. For $ \gamma \sim 1$, a single-component 
1D Bose fluid is a quasicondensate with a fluctuating phase but strongly 
suppressed density fluctuations~\cite{Petrov00Andersen02}. Large $\gamma$ 
leads to a crossover to the Tonks-Giradeau (TG) regime~\cite{Petrov00Andersen02}: 
the bosons become impenetrable, and  resemble  noninteracting fermions~\cite{Girardeau60}. 
The low-energy properties in both regimes can be well described by the Luttinger liquid
concept\cite{Haldane,Cazalilla02}.

These aforementioned regimes  
make up the state space of the  single-component 
system of 1D repulsive bosons. It may therefore seem
surprising that adding a second component to form a mixture of two types
of bosons ($B+B$) or of a boson plus a spin-polarized fermion ($B+F$)
can lead to a richer phase diagram. In this Letter, we show that such
binary mixtures are Luttinger liquids which can become unstable to 
gap-opening in one branch of the excitation spectrum, 
or to demixing, or to collapse.

These instabilities occur at low temperatures by tuning some parameters
of the system, signalling  a quantum phase transition (QPT)~\cite{Sachdev99}. 
Some of these instabilities
have direct analogues in higher dimensional systems, while others may be more
specific to 1D systems (and are analogues of 1D spin-$1/2$
fermionic systems~\cite{LutherEmery74,Gogolin98}). Dilute cold gases are ideal for
observing the instabilities that we predict because, in contrast to  conventional 
solid-state materials,  they are cleaner  and there is a large 
degree of control and tunability of parameters such as particle densities 
and the sign and strength of the interaction. 
Furthermore, diluteness makes it  {\it a priori} possible
to relate microscopic parameters such as scattering lengths and  atomic masses
to the  parameters of the low-energy effective theory.

  Contrary to their 3D counterparts~\cite{MathewsHall98,
TruscottSchreck01,mix,Timmermans98}, 1D binary  mixtures have only recently attracted  
interests~\cite{EckernParedes03,Das03}. To our knowledge, there has 
been no {\it analytic} treatment of these systems
within the Luttinger liquid framework. This is undertaken here using
the harmonic fluid approach, which  takes into account phase and density fluctuations in 1D 
fluids~\cite{Haldane,Cazalilla02}. We  summarize our results:

{\it Case 1.-}
We consider a binary mixture of two distinct atomic
species (either $B+B$ or $B+F$) or of two internal states of the same bosonic species. 
When a sufficiently repulsive interaction exists between the components, 
the mixture will be unstable against demixing, whereas for
sufficient attraction, it will collapse. 

{\it Case 2.-} We consider  a binary mixture  in {\it a low-density limit}, when 
the two components are sufficiently  close in density and if their sound velocities
are  similar, the system undergoes a QPT when there is an
attraction, no matter how weak, between the two components. The resulting state 
exhibits a strong pairing tendency. In the case of a $B+F$ mixture, the pairing fluctuations occur at  
the Fermi momentum, unlike  the $B+B$ case (or $F+F$~\cite{LutherEmery74}) where
they occur at zero momentum.

{\it Model and calculations:}  The 
experimental realization of the quasi one-dimensional systems requires 
a tight transverse confinement~\cite{Olshanii98,Petrov00Andersen02}, which nowadays
can be achieved in various setups~\cite{traps,Hansel01,Sauer01}.    
Our discussion below will focus on the properties of
homogeneous 1D systems. This should be appropriate for 
the central region of large harmonic~\cite{traps} or 
square-well traps~\cite{Hansel01}. Alternatively, a toroidal trap~\cite{Sauer01}
can provide the conditions for realizing a homogeneous 1D system. 
When necessary, we shall also discuss the 
effect of longitudinal confinement and finite size on the results to be presented below.

 Our  starting point will be the following Hamiltonian for a  mixture 
of two 1D dilute gases ($\alpha,\beta  = 1,2$): 
\begin{eqnarray}\label{ham}
H &=& \int dx \: \sum_{\alpha} \left[ \frac{\hbar^2}{2M_{\alpha}} 
\partial_{x}\Psi^{\dag}_{\alpha}(x)\partial_{x}\Psi_{\alpha}(x) - \mu_{\alpha} \rho_{\alpha}(x)\right] 
\nonumber\\
&+&  \frac{1}{2}  \int dx \: \sum_{\alpha,\beta} g_{\alpha\beta} \: \rho_{\alpha}(x) \rho_{\beta}(x).
\end{eqnarray}
For bosons, the field operators,
$\Psi_{\alpha}(x)$, obey  $\left[\Psi_{\alpha}(x),\Psi_{\alpha}^{\dag}(x') \right] = \delta(x-x')$, 
commuting otherwise.
For a spin-polarized fermion, the field operator obeys 
anti-commuting relations:
$\{\Psi_{2}(x), \Psi^{\dag}_{2}(x') \} = \delta(x-x')$. The density operators
$\rho_{\alpha}(x) = \Psi^{\dag}_{\alpha}(x) \Psi_{\alpha}(x)$. The inter-atomic interaction potential is 
$v_{\alpha \beta}(x) = g_{\alpha \beta} \delta(x)$, 
with $g_{\alpha\beta} = 2 \hbar \omega_{\perp} a_{\alpha\beta}$~\cite{Olshanii98}, where
$a_{\alpha\beta}$ are the  scattering lengths parametrizing the 3D interaction 
between species $\alpha$ and $\beta$, and $\omega_{\perp}$ is the transverse confinement frequency.

To study the low-energy properties of the above model, we use the
harmonic fluid approach~\cite{Haldane,Cazalilla02}, which 
employs the phase-density representation of the field operators. Phonon fields are introduced for 
each component: $\phi_{\alpha}(x)$ for long wavelength phase fluctuations
 and $\theta_{\alpha}(x)$ for long wavelength density fluctuations.
It is  assumed that  $\partial_x\theta_{\alpha}(x)$ and 
$\partial_x\phi_{\alpha}(x)$ are  small compared to the equilibrium densities $\rho_{0\alpha}$. 
This approach treats
bosons and fermions in 1D on equal footing~\cite{Haldane,Cazalilla02}, and  does not assume the 
existence of a BEC, as usual mean-field  treatments do. Therefore, it 
can describe both systems of impenetrable bosons 
(e.g. the Tonks gas~\cite{Girardeau60} and others~\cite{Cazalilla03}),  
and quasicondensates~\cite{Petrov00Andersen02}. Introducing~\cite{Haldane,Cazalilla02} 
$\Psi_{\alpha}(x) \sim 
\left[\rho_{0\alpha} + \partial_{x}\theta_{\alpha}(x)/\pi\right]^{1/2}\sum_{m} e^{im\pi \rho_{0\alpha}x 
+ im\theta_{\alpha}(x)} e^{i\phi_{\alpha}(x)}$, where the sum over $m$ involves only even (odd) integers
if the operator is bosonic  (fermionic), and $\rho_{\alpha} = \left[ \rho_{0\alpha} 
+ \partial_x \theta_{\alpha}(x)/\pi\right]
\sum_{m=-\infty}^{+\infty}  e^{2im\pi \rho_{{\rm o}\alpha}x + 2im\theta_{\alpha}(x)}$, 
into Eq.~(\ref{ham}) one obtains the low-energy Hamiltonian:
\begin{eqnarray}\label{hameff}
H_{\rm eff}= \frac{\hbar}{2\pi} \int dx \sum_{\alpha} \left[\frac{v_{\alpha}}{K_{\alpha}}
\left(\partial_x \theta_{\alpha} \right)^2 + v_{\alpha}K_{\alpha} 
 \left(\partial_{x} \phi_{\alpha} \right)^2\right] \nonumber\\ 
+ \frac{\hbar}{2\pi} \int dx \:\left[2 \tilde{g}_f\:  \partial_{x}\theta_1 \partial_x\theta_2 
+ \tilde{g}_{b} \: \cos 2(\theta_1 - \theta_2 +  \pi\delta x) \right].
\end{eqnarray}
In Eq.(\ref{hameff}),  we have retained only  
terms which can  have the most dominant effects at  low energies  
({\it i.e.}  marginal or  relevant in the renormalization group sense).
Its validity is restricted to energies smaller than the chemical potential.
The low-energy physics is fully characterized by the phenomenological  parameters $K_{1,2}$,
the sound velocities $v_{1,2}$,  the couplings $\tilde{g}_{\rm f,b}$, and $\delta = \rho_{01} - \rho_{02}$.
For small enough $g_{12}$, 
it is possible to relate these phenomenological parameters to the microscopic ones because
exchange and correlation effects will be of ${\rm O}(g^2_{12})$ in $\tilde{g}_{\rm f,b}$ (Otherwise, they must
be extracted from numerics or experiments~\cite{Cazalilla02}). First notice
that  $v_{N\alpha} = 1/(\hbar \pi \rho^{2}_{\rm o} \kappa_{S\alpha})$, where $\kappa_{S\alpha}  
= \rho_{\alpha}^{-2} (\partial\rho_{\alpha}/\partial\mu_{\alpha})$ is the 
compressibility~\cite{Haldane,Cazalilla02}. 
Furthermore, in homogeneous systems, Galilean invariance 
fixes the product $v_{J\alpha} = v_{\alpha} K_{\alpha} = v_{F\alpha} = 
\hbar \pi \rho_{0\alpha}/M_{\alpha}$~\cite{Haldane}. Thus
from the exact solution for a single-component 1D Bose fluid~\cite{Lieb63}, 
$v_{N\alpha} =  v_{\alpha} K^{-1}_{\alpha}  = v_{F\alpha} \left[1 - 8 \gamma_{\alpha}^{-1} 
+ O(\gamma^{-2}_{\alpha})\right]$ for $\gamma_{\alpha} = M_{\alpha} 
g_{\alpha\alpha}/\hbar^2 \rho_{0\alpha} \gg 1$ and 
$v_{N\alpha} \simeq v_{F\alpha} \gamma_{\alpha}/\pi^2$, for $\gamma_{\alpha} \lesssim 1$. 
Therefore,  $1\leq K_{\alpha} < +\infty$ for bosons, $K_{\alpha} = 1$
in the Tonks limit. For spin-polarized fermions, $K_{2} \simeq 1$, because
the s-wave scattering length vanishes thanks to  the Pauli principle, leaving  the much weaker 
p-wave channel, which can be neglected.
Finally, in the weak coupling limit,  $\tilde{g}_{f} \simeq g_{12}/(\hbar\pi)$ 
and  $\tilde{g}_{b} \simeq 4\pi g_{12} \rho_{01}\rho_{02}/\hbar$.
The sign and strength of $g_{12}$ can be controlled using a Feshbach  or a confinement 
induced~\cite{Olshanii98} resonance. Interestingly, the same Hamitonian,
Eq~(\ref{hameff}), also describes a binary mixture  in a 1D optical lattice~\cite{unpub} 
provided that none of the components is commensurate with the lattice periodicity. 
This allows for further  possibilities of tuning the parameters 
$v_{\alpha}$, $K_{\alpha}$, and $g_{12}$ (e.g. the strength of $g_{12}$ can be modified
by shifting relative to each other the lattices where each component hops).

The above Hamiltonian, Eq.~(\ref{hameff}), describes a rich variety of one-dimensional binary mixtures:

{\it Case 1.-}  When the equilibrium densities of the two species are different, i.e. for sufficiently large 
$\delta = \rho_{01} -\rho_{02}$,   the cosine term in Eq.~(\ref{hameff}) can be neglected (see below).
Thus one is left with two Luttinger liquids  
coupled by   $\tilde{g}_{f}\: \partial_x \theta_1 \partial_x \theta_2$. The normal
modes of the system can be  found from the equations of motion for $\theta_1(x,t)$ 
and $\theta_2(x,t)$.   The phase velocity of the normal modes is:
\begin{equation}\label{veloc}
v^2_{\pm} = \frac{1}{2}(v^2_1 + v^2_2) \pm \frac{1}{2}\sqrt{\left(v^2_1 - v^2_2\right)^2 
+ 4 (v_{J1}) (v_{J2}) \tilde{g}^2_{f} }.
\end{equation}
Hence, the mixture will become unstable provided that
\begin{equation}\label{instab}
\sqrt{ v_{N1} v_{N2} }  < |\tilde{g}_{f}|.
\end{equation}
For a repulsive interaction ($\tilde{g}_{f} > 0$),
the two  components of  the mixture repel each other sufficiently strongly to demix. 
However, for  an attractive interaction
($\tilde{g}_{f} < 0$), the system is expected to collapse.

  We now examine the linear instability condition, Eq.~(\ref{instab}), in several limiting cases. 
For a $B+B$ mixture in the quasicondensate regime ($\gamma_{1,2} \lesssim 1$),
$v_{N\alpha} \simeq v_{F\alpha} \gamma_{\alpha}/\pi^2$ and $\tilde{g}_{f}\simeq g_{12}/(\hbar\pi)$.
Then the mixture is unstable when $\sqrt{g_{11} g_{22}} < |g_{12}|$. In higher dimensional
systems this condition is obtained from a mean-field theory~\cite{Timmermans98}, which implicitly assumes the
existence of a condensate. It is therefore interesting that it also applies to 1D systems in
the quasicondensate regime, where the phase fluctuates. Another interesting limit is 
a $B+F$ mixture where the boson is a quasi condensate. For the fermion, 
$v_{N2} \simeq v_{F2} = \hbar \pi \rho_{02}/M_2$. Hence,
the system will become unstable  below
a critical fermion density $\rho^{\rm crit}_{02} = M_2 g^2_{12}/(\pi^2 \hbar^2 g_{11})$.
This agrees with the mean-field result of Das~\cite{Das03}.
Lastly,  consider a  $B+F$ or a $B+B$ mixture with impenetrable boson(s)
($\gamma_{\alpha} \gg 1$). Then Eq.~(\ref{instab}) implies that the 
system becomes unstable when   
$\left(\rho_{01}\rho_{02}\right) <   M_{1}M_{2} g^2_{12}/(\pi^4 \hbar^4)$. 

	We now  briefly discuss the properties of a stable mixture.
As  found above, at long wavelengths there will be two branches of modes with
phase velocities $v_{\pm}$ given by Eq.~(\ref{veloc}). This result is not qualitatively affected by
the presence of a longitudinal confinement, though the actual energies of the modes may. More interestingly, 
a stable binary mixture will be a Luttinger liquid with all correlations characterized  (in the thermodynamic limit
and at zero temperature) by the usual power laws~\cite{Haldane}. In a trapped system, the power laws will
be accurate near the center of the trap~\cite{Cazalilla02}. A full discussion of these issues will be 
given elsewhere~\cite{unpub}. 

{\it Case 2}: When the components of the mixture are 
close enough in density,  one can no longer neglect 
the cosine term in Eq.~(\ref{hameff}). To see this, consider
the Tonks limit where the bosons  behave as free fermions. Then this
term describes a backscattering process between the fermions: A fermion of 
each species is excited from one Fermi point to the other, exchanging a momentum 
close to $2 \pi (\rho_{0} \pm \delta)$. If $\delta \ll \rho_0 =  (\rho_{01}+\rho_{02})/2$ 
then this process can occur at low energies and needs to be taken
into account. To simplify our analysis, we take  $\delta = 0$  ($\delta \neq 0$ will 
be addressed at the end). By a perturbative analysis of the free energy, 
we see that the  correction due to this term becomes larger than the zeroth order
term as $T\to 0$ for $K_1 + K_2 < 2$.  To go further, we use 
the renormalization group (RG) method and 
take $K_1 = K_2 = 1$  and  $v_1 = v_2$ (We have checked that
small deviations from these assumptions do not alter the 
physical picture qualitatively~\cite{unpub}).  The rationale for
this   is that   the lower bound in  a boson system is $K _{\alpha} = 1$, which is reached
for the Tonks gas limit (i.e. by reducing the density, increasing the scattering lengths  $a_{\alpha\alpha}$ or both~\cite{Petrov00Andersen02}).  In a $B+F$ mixture, however,  the fermion already has $K_2 \simeq1$, 
and we only need to tune the boson  to the Tonks   limit. 
Furthermore, if the velocites $v_1$ and $v_2$ can be made made similar~
\footnote{On the continuum, this is the case for a $B+B$ mixture of two internal states of $^{87}$Rb~\cite{MathewsHall98} 
or a $B+F$ mixture of two isotopes of Li~\cite{TruscottSchreck01} because their masses are equal or similar. 
On an optical lattice, one can adjust the lattice parameters to have $v_1 \simeq v_2$ even if 
$M_1 \neq M_2$.}, then
in the Tonks limit at equal densities, 
$v_1  = v_2 = v_F$. Thus the Hamiltonian~(\ref{hameff}) becomes
$H_{\rm eff} = H_{+} + H_{-}$, where:
\begin{eqnarray}
H_{+} &=& \frac{\hbar v_{+}}{2\pi} \int dx\:  \left[ K^{-1}_{+}\left(\partial_x\theta_{+} \right)^2 + 
K_{+}\left(\partial_{x}\phi_{+}  \right)^2  \right],\\
H_{-} &=& \frac{\hbar v_F}{2\pi} \int dx\:  \left[ \left(\partial_x\theta_{-} \right)^2 + 
\left(\partial_{x}\phi_{-}  \right)^2  \right]   \nonumber \\
&+&\frac{\hbar}{2\pi}\int dx\: \left[  \tilde{g}_b \cos{\sqrt{8}\theta_{-}(x)} -\tilde{g}_f\left(\partial_{x}\theta_{-}\right)^2\right].
\end{eqnarray}
The new fields $\theta_{\pm}(x) = \left(\theta_1(x) \pm \theta_2(x)\right)/\sqrt{2}$ and 
$\phi_{\pm}(x) = \left(\phi_1(x) \pm \phi_2(x)\right)/\sqrt{2}$ describe the in-phase (``$+$") and 
out-of-phase (``$-$") density and phase fluctuations of the components of the mixture.
The fact that in this limit they are decoupled is an analog of spin-charge separation,
so ubiquitous in 1D Fermi systems~\cite{Gogolin98,Recati03}. 

  $H_{+}$  has the form of a Luttinger liquid Hamiltonian characterized by  the parameters 
$K_{+} = \left[1 + \tilde{g}_f/v_F \right]^{-1/2} \simeq \left[ 1 + g_{12}/(\hbar\pi v_F)  \right]^{-1/2}$ 
and $v_{+} = v_F\left[1 +  \tilde{g}_f/v_F \right]^{1/2} \simeq v_F\left[1 +  g_{12}/(\hbar \pi v_F) \right]^{1/2}$.
Hence, in-phase modes are gapless (in the thermodynamic limit).

$H_{-}$ is a sine-Gordon model~\cite{Gogolin98}.
The relative importance  of the marginal operators, the cosine term and
$(\partial_x\theta_{-})^2$, can be assessed using  the 
RG at weak coupling, where 
$g_f = \tilde{g}_f/v_F \ll 1$  and $g_b = 
 \rho^{-2}_{0} \tilde{g}_b/4 \pi^2 v_F \ll 1$.   
To second order in these couplings, the scaling flow is
of Berezinskii-Kosterlitz-Thouless (BKT) type
(see e.g. Refs.~\cite{Gogolin98,Sachdev99}). It is interesting to note that
if the sound velocities are tuned to be equal (i.e. $v_1 = v_2$) in the limit $K_1 = K_2 = 1$, 
then the RG flow proceeds along the separatrix $g = g_f = g_{b}\propto g_{12}$
and the spectrum of the system exhibits an enhanced  SU$(2)$ symmetry. This
is not present in the microscopic Hamiltonian~(\ref{ham}), and seems quite
striking, especially for a $B+F$ mixture.  Thus,
for small $g_{12} > 0$, the {\it effective} coupling $g$
is renormalized to zero as the temperature is decreased and 
the cosine term  leads only to subleading 
corrections to the low-temperature properties: 
the out-of-phase modes will behave as a Luttinger liquid with
$K_{-} =  1$ and $v_{-}  = v_F$.  
But when small $g_{12} < 0$,  the 
{\it effective} coupling $g_b\to + \infty$ with decreasing  temperature. 
The exact solution of the sine-Gordon model~\cite{Gogolin98} then 
tells us that a gap opens for the excitations of the out-of-phase mode.  
At weak $g \propto g_{12}$ the gap $\Delta \sim |g|^{1/2}\: e^{-1/|g|}$~\cite{Gogolin98,Sachdev99},
i.e. exponentially small, and  it becomes larger  as $g_{12}$ increases. Hence,
at low enough temperatures only in-phase modes can be excited. In a finite-size 
system, however, the RG flow to strong coupling will be cut-off at a length scale comparable to the
size of the system. Therefore, for weak attraction one should observe that the out-of-phase modes
become more ``stiff'' until the full gap develops.

If  $v_1\neq v_2$ and/or $K_1, K_2 \neq 1$ (with
the differences small) the effective SU(2) symmetry is lost  and the 
transition is of BKT type, the gap  again being exponentially small but with a different
functional form (see e.g.~\cite{Gogolin98,Sachdev99}).   One notable consequence
of the development of the gap is that density fluctuations of both components will become
correlated because only in-phase modes are allowed at low enough temperatures:
specifically, the $2\pi \rho_{0}$ part of density correlation
 $\langle \rho_{\alpha}(x) \rho_{\beta}(0) \rangle\big|_{2\pi\rho_{0}}
\sim \cos(2\pi \rho_{0})\: x^{-K_{+}}$ for arbitrary $\alpha = \beta = 1,2$ {\it and} 
$\alpha\neq \beta$. These correlations can be observed by exciting the modes 
of the system and imaging each species independently.

 The aforementioned correlations are also reflected in  the ``pairing'' correlations described
 by the operator $\Delta(x) = \Psi_1(x)\Psi_2(x)$.  
For a $B+B$ mixture, $\langle \Delta^{\dag}(x) \Delta(0)\rangle \sim x^{-1/K_{+}}$, which 
decays more slowly than the  density correlation, since for $g_{12} < 0$, $K_{+}  \simeq 
\left[1  + g_{12}/(\hbar \pi v_F)\right]^{-1/2} > 1$. For a $B+F$ mixture,  the
same correlation function has an interesting oscillation at $\pi \rho_0$ (rather than at
$2 \pi \rho_0$): 
$\langle \Delta^{\dag}(x) \Delta(0)\rangle \sim \sin\left(\pi\rho_{0} x\right)\:  x^{-1/K_{+} - K_{+}/4}$. 
Thus, in a $B+F$ mixture, the pairing correlations will decay   
more slowly than the density correlations for $K_{+} > 2/ \sqrt{3} \simeq 1.16$. 

\begin{figure}[t]
\includegraphics[width=\columnwidth]{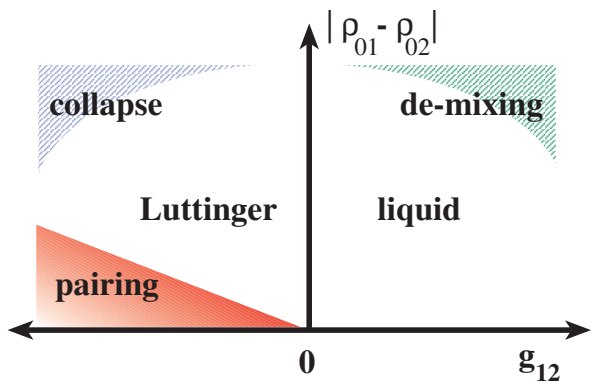}
\caption{Schematic phase diagram  with the boson(s) in the 
Tonks limit. $|\rho_{01}-\rho_{02}|$ is the density difference at $g_{12}=0$.}\label{fig}
\end{figure}

 We now discuss the physical picture underlying the above calculations.
On intuitive grounds, an attraction between the two components of the mixture ($g_{12} <0$)
will lead to the formation of bound pairs, and might eventually lead to collapse. 
However, in the regime studied  above, pairing takes place between 
hard-core bosons or a hard-core boson and a fermion, which means that
as long as the repulsion between bosons of the same species is stronger, only bound pairs containing
one particle of each component are possible. This picture is confirmed by considering a 
strong-coupling limit of the Hamiltonian~(\ref{ham}), where 
first we take $g_{11},g_{22} \to +\infty$ and then $g_{12}\to -\infty$ such that 
$|g_{12}| \ll g_{11},g_{22}$~\footnote{For a $B+B$ mixture, the limit of $g_{11}, g_{22}\to +\infty$ 
but finite $g_{12}$ maps to an integrable model of spinful fermions solved by
C.~N. Yang. For $g_{12} <0$ the fermions form bound pairs.
See e.g. M. Takahashi, {\it Thermodynamics of One-Dimensional Solvable Models}, 
Cambridge University Press (Cambridge, 1999). }. 
Moreover, in a $B+B$ mixture pairs are also bosons, 
but their condensation is forbidden in 1D dimension by rigorous theorems~\cite{Petrov00Andersen02}. In a 
$B+F$ mixture, the pairs are fermions, and this slightly suppresses pairing correlations relative to the $B+B$ case. 
We expect a smooth crossover from the weak coupling regime  to the strong-coupling regime 
where tightly bound pairs form. As the strength of attraction between
the two components increases, the size of the gap also grows thus
preventing collapse (Fig.~\ref{fig}).

 However, the gap can be destroyed when a sufficiently large difference in chemical potential
leads to a density difference $|\delta| > 0$. 
This transition is of  commensurate-incommensurate type~\cite{Gogolin98}. 
Small inhomogeneities in the density due to the trap, will not have a strong effect in the pairing 
provided that the density difference  between the two components 
does not become too large. Otherwise, one can expect phase segregation into pairing 
and nonpairing regions, with the former occurring mainly near the center of the trap.

In conclusion, we have shown that binary mixtures of 1D quantum gases exhibit 
some interesting instabilities. In particular, we have found a pairing instability which may be special
to one dimension and should have unusual experimental signatures~\cite{unpub}. 

We acknowledge inspiring conversations with A. Nersesyan and M. Fabrizio, and useful 
correspondence with B. Svistunov. A.~F.~H. thanks the hospitality of 
DIPC and a grant from EPSRC (U.K.). M.~A.~C. thanks  Gipuzkoako Foru 
Aldundia for finacial support.

\end{document}